\documentclass[12pt,landscape,preprint]{aastex}

\usepackage{url}
\usepackage{epstopdf}
\newcommand {\gtsim} {\ {\raise-.5ex\hbox{$\buildrel>\over\sim$}}\ }
\newcommand {\ltsim} {\ {\raise-.5ex\hbox{$\buildrel<\over\sim$}}\ }

\shorttitle{Pseudobulges in NGC 5746}
\shortauthors{Barentine and Kormendy}

\begin{document}

\title{Two Pseudobulges in the ``Boxy Bulge'' Galaxy NGC 5746}   
\author{John C. Barentine} \author{John Kormendy}
\affil{Department of Astronomy\\
University of Texas at Austin\\ 
1 University Station C1400 \\
Austin, TX 78712-0259, USA}    

\begin{abstract}
Galaxy formation and growth under the $\Lambda$CDM paradigm is expected to proceed in a hierarchical, 
bottom-up fashion by which small galaxies grow into large galaxies; this mechanism leaves behind large 
``classical bulges'' kinematically distinct from ``pseudobulges'' grown by internal, secular processes.  We use archival data (\it Spitzer \rm 3.6 $\mu$m wavelength,  \it Hubble Space Telescope \rm $H$-band, Two Micron All Sky Survey $K_s$-band, and Sloan Digital Sky Survey $gri$-band) to measure composite minor- and major-axis surface brightness profiles of the almost-edge-on spiral galaxy NGC 5746.  These light profiles span a large range of radii and surface brightnesses to reveal an inner, high surface brightness stellar component that is distinct from the well-known boxy bulge.  It is well fitted by S\'ersic  functions with indices $n = 0.99 \pm 0.08$ and  1.17 $\pm$ 0.24 along the minor and major axes, respectively.  Since $n < 2$, we conclude that this innermost component  is a secularly-evolved pseudobulge that is distinct from the boxy pseudobulge.  This inner pseduobulge makes up 0.136 $\pm$ 0.019 of the total light of the galaxy.  It is therefore considerably less luminous than the boxy structure, which is now understood to be a bar seen nearly end-on.  The infrared imagery shows further evidence for secular evolution in the form of a bright inner ring of inner radius 9.1 kpc and width 1.6 kpc.  NGC 5746 is therefore a giant, pure-disk SB(r)bc galaxy with no sign of a merger-built bulge.  We do not understand how such galaxies form in a $ \Lambda$CDM universe.
\end{abstract}

\keywords{galaxies: individual (NGC 5746) --- galaxies: bulges --- galaxies: evolution}

\section{Introduction\label{sec:intro}}

\pretolerance=15000  \tolerance=15000

The $\Lambda$CDM paradigm is based on the observation of dark matter concentrations (``halos'') into 
which baryons fall, cool, and form stars.  Galaxies are built by the hierarchical merging 
of smaller halos in a bottom-up fashion \citep{WhiteRees78}.  This implies frequent ``major'' mergers 
in the early Universe between objects of comparable mass.  Mergers are understood to scramble disks
into elliptical galaxies \citep{Toomre77, Schweizer90} and are accompanied by cold-gas 
dissipation and starbursts in the smaller galaxies but not in the biggest galaxies that can hold onto X-ray gas 
(see \citealt{KFCB} for a review).    The rate of major mergers peaked at $z$ $\sim$ 1.2 \citep{mergerpeak}, and 
since that time, the Universe has been in transition from merger-dominated galaxy evolution to slow
(``secular'') internal evolution \citep{KFCB}.  

In spite of the large amount of merger violence at early times, almost 2/3 of the most massive galaxies 
in the nearby universe (field environments like the Local Group) show no classical bulge at all 
(\citealt{Kormendy10, FisherDrory11}; see
\citealt{Peebles10}
for a review).
Thus most field galaxies show no sign that they experienced a major merger at least 
since the time of the merger rate peak.
It is difficult to understand how stellar disks that were already in place at
$z \sim 1$ survived until today without being converted into -- or at least augmented by -- a classical
bulge \citep{Kormendy10}.  Here, it is important to note that we do not have the freedom to postulate
bulges whose properties make them easy to hide.  Bulges satisfy well defined ``fundamental plane''
parameter correlations (see 
\citealt{KB12}
for the most recent version).  Low-luminosity classical bulges are small, but they have high surface
brightnesses and are described by S\'ersic functions  \citep{Sersic68} with indices $n \simeq 2.5$ that make 
them hard to hide and easy to identify.  Thus we can be confident that, while $\Lambda$CDM performs well on 
large scales, the large fraction of nearby, giant galaxies which show little evidence of major mergers presents a 
challenge to our understanding of galaxy evolution on 10-kpc scales (e.{\thinspace}g., \citealt {Abadi03, 
Governato10, Kormendy10, Peebles10}).

The magnitude of this challenge is underestimated if effectively bulgeless galaxies are 
undercounted because pseudobulges are mistaken for classical bulges.  Prior to the development of
the secular evolution picture (especially \citealt{Kormendy93,KFCB}), this routinely happened
even for face-on galaxies with dynamically disky pseudobulges (e.{\thinspace}g., NGC 4736; see the above
references).  But identification of classical bulges in edge-on galaxies is particularly tricky.
``Box-shaped bulges'' such as that in NGC 4565 \citep{KB10} are made of old stars and clearly bulge out above 
and below the edge-on disk plane.  It is understandable that, in early papers, they were thought to be classical 
bulges which happen to have a box-shaped peculiarity in their structure.  But now we know that box-shaped 
bulges are galaxy bars that are seen edge-on (\citealt{CombesSanders81, Combes90, PfennigerNorman90, 
PfennigerFriedli91, Raha91, AthanassoulaMisiriotis02, Athanassoula05, Shen10}).  Bars form by gravitational 
instabilities in galaxy disks.   The above papers show that, once a bar is well formed, it buckles vertically and 
rapidly turns into a thick structure that looks like a \hbox{box-shaped} bulge when seen \hbox{edge-on.}  Heating 
of stars whose vertical motions are in resonance with the bar density wave further contributes to the thick, boxy-
distorted structure.  Also, a splitting of gas rotation velocities in edge-on boxy bulges (a ``figure 8'' shape of 
spectral emission lines) is a robust signature of gas flow in an edge-on bar and further cements our view of these 
boxy structures as bars (\citealt{Kuijken95, Merrifield96, MerrifieldKuijken99, BureauFreeman99}).  Thus, a 
``boxy bulge'' in an edge-on object would be identified as the galaxy's bar if it were seen more face-on.  Since it 
is really a part of the disk, we call such structures ``boxy pseudobulges''.  This means that the true classical-
bulge-to-total luminosity ratios of galaxies such as NGC 4565 are much smaller than the values $\sim 1/3$ that 
are measured (e.{\thinspace}g., \citealt{SimienGdV86}) when the boxy structure is identified as the bulge.

We demonstrated this effect for the nearly edge-on spiral NGC 4565 in \cite{KB10}.   Its center has 
a boxy photometric signature previously identified as a classical bulge \citep{SimienGdV86}.
But we found that this component is well fitted by a S\'{e}rsic function with 
index $n$ = 1.  This demonstrates that it does not have the characteristics of a merger-built 
classical bulge \citep{KFCB,FisherDrory08}.  It is consistent with a bar seen nearly end-on.
Mid-infrared observations penetrate the thick midplane dust, revealing the true central component
in this galaxy.  It is a pseudobulge whose scale height is smaller than that of the outer disk.  
Its small pseudobulge-to-total ($PB/T$) ratio, 0.061$^{+0.009}_{-0.008}$, means that the disk and its 
secularly-built structures completely dominate this galaxy.  NGC 4565 is a massive galaxy; its rotational 
speed is $\sim$255 km s$^{-1}$ interior to its outer warp \citep{Rupen91}.  This result is especially 
hard to understand because it is easier to make bulgeless galaxies via feedback mechanisms \citep
{Governato10} when the resulting galaxy is a dwarf.    Thus NGC 4565 and galaxies like it are problematic in the 
context of hierarchical assembly models.
 
We know that galaxies like NGC 4565 are not rare \citep{Kormendy10,FisherDrory11}.  Previous efforts to 
measure $B/T$ ratios of galaxies by way of light profile decompositions may have resulted in overestimates
 of the bulge contribution in galaxies at moderate to high inclinations.  This could mask the true number 
of essentially bulgeless galaxies in the Universe.  The problem of apparently bulgeless galaxies becomes 
more acute because most decomposition work is now done in an automated fashion and in two dimensions.
It is difficult for these analyses to cope with patchy internal obscuration.
Here, we use carefully constructed one-dimensional light profiles to address this problem.

The aim of this paper is to determine for NGC 5746, i.{\thinspace}e., an additional normal, edge-on disk galaxy 
with a boxy bulge, the fraction of the total galaxy luminosity that is contributed by the boxy structure and by any 
additional, disky pseudobulge near the center.  This requires making a clear photometric distinction between 
structural components in circumstances compromised by strong dust absorption.  To this end, we determine the 
pseudobulge-to-total luminosity ratio $PB/T$ and the pseudobulge scale height.  In \S2 we describe the 
selection of this galaxy and outline the method by which we constructed minor- and major-axis light profiles from 
photometric data spanning a range of wavelengths from the optical to the mid-infrared.  We present the profiles 
in \S3 and compare them with previous studies.  The galaxy's $PB/T$ ratio is calculated using these profiles, 
after subtracting fits to the other structures seen in the profiles.  We show that NGC 5746 is a massive disk galaxy 
in which the bulge does not dominate the 
light profile at any radius. We summarize our results in \S4.

\section{Method\label{method}}

\subsection{Target Selection}

We searched various lists of edge-on galaxies and applied the following selection rules: candidate objects must 
(1) be nearby (distance $D \lesssim$ 75 Mpc); (2) be highly inclined (inclination $i \gtrsim 85^{\circ}$); (3) show 
minimal, if any, indication of recent interaction with other galaxies; (4) have available data over a wide range of 
wavelengths and resolutions; and (5) be relatively free of dust.  NGC 5746 ($\alpha_{2000.0}$=14$^{h}$44$^
{m}$56$\fs$005, $\delta_{2000.0}$=+01$^{\circ}$57$\arcmin$17$\farcs$06) meets essentially all of these 
requirements.  It is classified as an SAB(rs)b? in the Third Reference Catalog of Bright Galaxies (RC3; \citealt
{RC3}).  Measurements reported in the literature give a mean distance of 27.6 Mpc with a 1$\sigma$ dispersion 
of 2.5 Mpc (\citealt{NBC}; \citealt{Willick97}; \citealt{Rand08}; \citealt{Tully_etal08}; \citealt{Springob09}; \citealt
{ErrSpringob}; \citealt{Tully_etal09}).  The galaxy is inclined to the line of sight by 83.9$^{\circ}$ and has a rather 
large maximum circular velocity of 318.5 $\pm$ 9.8 km s$^{-1}$.  The inclination and velocity were obtained from 
HyperLEDA\footnote{The Lyon-Meudon Extragalactic Database; \url{http://leda.univ-lyon1.fr/}} and the circular 
velocity is corrected for inclination.  While not as highly inclined as we would prefer, the observed angle has the 
benefit of reducing the effect of the dust in the central region.  This galaxy also has the largest amount of archival 
photometric data of the candidate objects we considered.

NGC 5746 has been shown previously to contain kinematic evidence for the presence of a bar \citep
{Kuijken95,BureauFreeman99}, manifesting itself in the apparent box-like shape of the ``bulge'' in optical 
images.  We confirm the existence of the bar in observations reported here.  The question remains: if the 
apparent bulge is actually the photometric signature of the bar, then where is the bulge in this galaxy?

\subsection{Data and Calibrations\label{subsec:method}}

In many edge-on disk galaxies other than S0's, extinction at optical wavelengths is very large along sightlines 
through the disk midplane and could hide structures with scale heights smaller than that of the dust.  
Observations in the near- and mid-infrared can help overcome this problem.  To see through the dust, we used 
3.6 $\mu$m images made with the Infrared Array Camera (IRAC; \citealt{Fazio04}) aboard the \it Spitzer Space 
Telescope \rm to measure the minor- and major-axis light profiles NGC 5746.  The spatial resolution of IRAC is 
insufficient to extend the light profiles to the smallest radii, so in the innermost  region we augmented the profile 
with $H$-band data from the \it Hubble Space Telescope \rm  Near Infrared Camera Multi Object Spectrometer 
(NICMOS; \citealt{NICMOS}).  At large radii, where dust is less of a problem and maximizing the signal-to-noise 
ratio ($S/N$) becomes more important, we used data from two large  sky surveys: the Two Micron All-Sky Survey 
(2MASS; \citealt{2MASS}) and the Sloan Digital Sky Survey (SDSS; \citealt{YorkSDSS}).  Filters, pixel scales, 
and fields of view of each telescope and instrument are given in Table~\ref{instruments}.   We briefly summarize 
the calibration procedures for each data source. 

The IRAC data were reduced by the Spitzer Science Center (SSC) using software pipeline version S14.0.0.  The 
reduction steps include subtracting the bias level and dark current, flat fielding, and performing sky subtraction.  
We used the final, mosaicked versions of the images containing all pointings of the telescope at a given location 
and time of observation.  

NICMOS data were calibrated using version 4.4.0 of the CALNIC reduction pipeline \citep{CALNIC}.  The code 
applies basic corrections to the data, including dark current subtraction, corrections for detector non-linearity, 
and flat fielding.  After all images in an ``association'' of data are processed in this manner, a second stage 
creates mosaics of overlapping images and subtracts a scalar background (``sky'') value.  However,  proper sky 
subtraction of images of extended sources is severely impacted by the instrument's small field of view.  NGC 
5746 overfills the NICMOS frame such that at no location is the true sky level reached; the automated data 
reduction procedures typically overcorrect for sky by subtracting a value higher than the true sky level, resulting 
in negative pixel values in the corners of frames.  We accounted for this by locating NICMOS frames from other 
programs taken as close in time as possible to our galaxy observations, typically within one day, and measuring 
real sky values from frames that did not contain large, extended objects.  The ``sky'' value recorded in the 
headers of our galaxy images removed by CALNIC is added back to the pixels in our images and the measured 
sky value subtracted off.  

The 2MASS data were calibrated nightly during survey operations by observing standard star fields at regular 
intervals.  Photometry of the standard stars was used to derive the extinction coefficients and photometric 
zeropoints in each of the three survey passbands as a function of time throughout a given night.  The survey 
observations did not permit absolute calibration of the 2MASS photometric system.  However, \citet{2masscalib} 
offer a calibration tied to Vega on the ``Cohen-Walker-Witteborn'' system from computation of relative spectral 
response curves.  To place our fluxes on the 2MASS system, we used the zeropoints computed by the reduction 
pipeline and written into the image headers.

Photometric data from SDSS were calibrated using the PHOTO pipeline \citep{PHOTO} which gathers 
astrometric data, obtains the extinction and photometric zeropoint on the night the data were collected, renders 
the drift-scan images into a series of postage stamps, and estimates the flat field vectors, bias drift, and the sky 
level for each.  Corrected frames are produced using this information.  We did not make use of SDSS-generated 
PSF or Petrosian-fitting photometry, but rather performed our own surface photometry on corrected frames.

\subsection{Surface Photometry\label{subsec:photometry}}

Images from a given source were first prepared by correcting for any background gradients and cleaning 
contaminating pixels from sample regions.  Systematic variations in the background were removed with the IRAF
\footnote{IRAF is distributed by the National Optical Astronomy Observatory, which is operated by the 
Association of Universities for Research in Astronomy (AURA) under cooperative agreement with the National 
Science Foundation.} task \tt mscskysub\rm, which fits a polynomial or spline of arbitrary order to a 2D surface.  
We specifically used second-order polynomials.  Contaminants consisting of cosmic rays, foreground stars, and 
background objects were removed by computing the median of pixels in a user-defined region around the 
source; pixels varying from the median by more than 1.5$\sigma$ were replaced by the median.  This cleaning 
was done carefully by hand, and individual, problematic pixels were replaced on a case-by-case basis.  
Extended objects were treated by interpolating over pixels in a user-defined box around an object with a 
second-order polynomial and replacing the pixel values in the box with those of the fit.  This approach works well 
for moderately bright stars but breaks down for the brightest objects; the affected pixels were edited out of the 
extracted 1D light profiles by hand.

\begin{figure}
\plotone{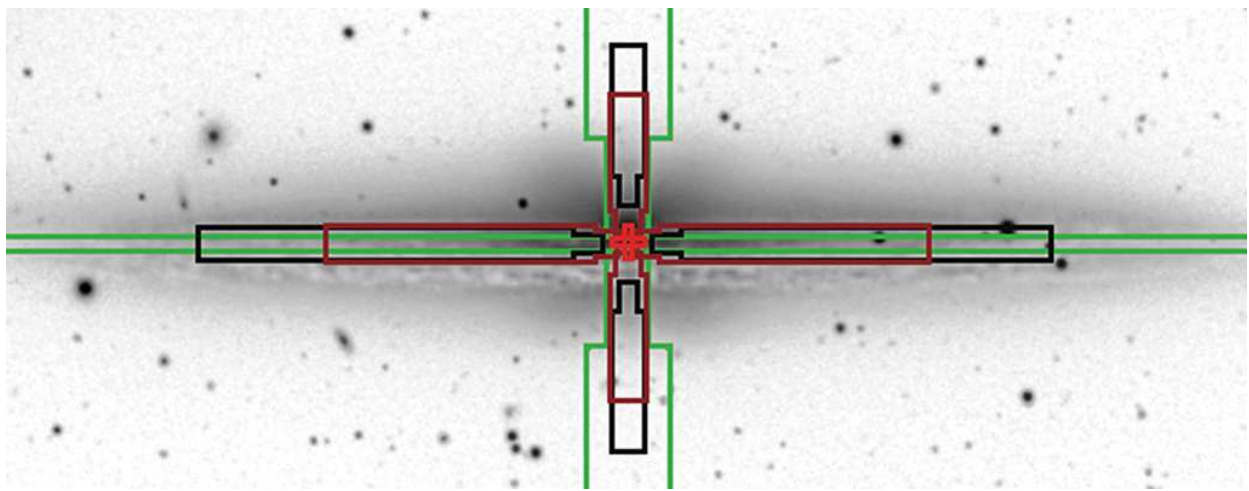}
\caption{An 8100 s $R$-band image of NGC 5746, rendered with an inverted colormap, onto which has been 
superimposed a set of colored overlays representing the effective sizes and shapes of the cut boxes used in 
performing surface photometry.  The image has been rotated such that the major axis is aligned with the image 
rows.  The colors of the boxes indicate the data sources: SDSS (green), HST NICMOS F160W (red), 
2MASS $K_s$ (brown), and Spitzer IRAC 3.6 $\mu$m (black), and match the colors of the data points in the 
minor- and major-axis light profiles presented in Figures~\ref{minor_profile} and \ref {major_profile}.  The 
radial extent of the boxes reflects the radial range of points plotted in the light profiles.}
\label{cutboxes}
\end{figure}

We performed surface photometry on NGC 5746 by taking rectangular cuts along its minor and major axes, 
supplemented where possible with ellipse fits of isophotes that served as a check on the cuts.  We generally 
avoided ellipse fitting in favor of cuts because the isophotes are far from elliptical over a large range of radii, and 
ellipse fits tend to fail in the presence of significant midplane dust absorption.  For each data source, we 
investigated a range of cut widths over different radius ranges in order to construct light profiles of the highest 
possible signal-to-noise ratio ($S$/$N$) at large radii while preserving resolution at small radii.  Consequently, 
the cut boxes effectively had stairstep shapes tapering to progressively narrower widths at smaller radii; the effective
dimensions and orientations of these cut boxes are illustrated in Figure~\ref{cutboxes}, in which they are
superimposed over an optical image of the galaxy.  The sky level in each image was determined by sampling 
regions at large radii as free from contamination by foreground 
and background objects as possible. 1D profiles were extracted by block-averaging pixels along the short 
dimension of a cut box in IRAF.   In order to mitigate the effect of midplane dust absorption where it interferes 
with the minor-axis light profile, we excised points by hand from the profile that were obviously impacted 
adversely by the dust.  For the major-axis profile, we defined the cut box parameters to carefully avoid the dust 
lane (see Figure~\ref{cutboxes} for placement).

All photometric data were given an absolute calibration by tying them to the 2MASS $K_s$ points in order to 
place them on a common photometric system.  To assemble a composite light profile from a number of data 
sources, arbitrary constants were added to light profiles generated for each source individually to bring data 
points into coincidence with those of the 2MASS profile over the radius regime in which they overlap.   Profiles 
tend to plateau at small radii for data sources with poor spatial resolution; these points were trimmed from the 
final profile of each object.  Additionally, points were trimmed at large radii at which the instrumental sensitivity 
rapidly diminishes, otherwise leading to an underestimate of the surface brightness.  We present the final 
versions of the light profiles for each object with symbol colors corresponding relatively to the wavelengths of the 
data sources to help guide the eye.

\subsection{1D Radial Profile Decompositon\label{subsec:profiles}}

After extracting the profiles, we decomposed them into three components: a S\'{e}rsic function for the central 
pseudobulge, another S\'{e}rsic function for the boxy bulge, and an outer exponential representing the disk.  The 
choice of an exponential is consistent with the \citet{vdKruit_Searle_81_1} model of the disk as a locally 
isothermal sheet.  All components were fitted with a ${\chi}^2$-minimization algorithm employing the simplex 
optimization method.  Our fitting code allows for the simultaneous decomposition of a given profile into a single S
\'ersic and single exponential function with five free parameters: the radius and surface brightness of the bulge 
($R{_n}, {\mu}{_n}$), the radius and surface brightness of the disk ($R{_d}, {\mu}{_d}$), and the S\'ersic index, $n
$.   

\begin{figure}
\plotone{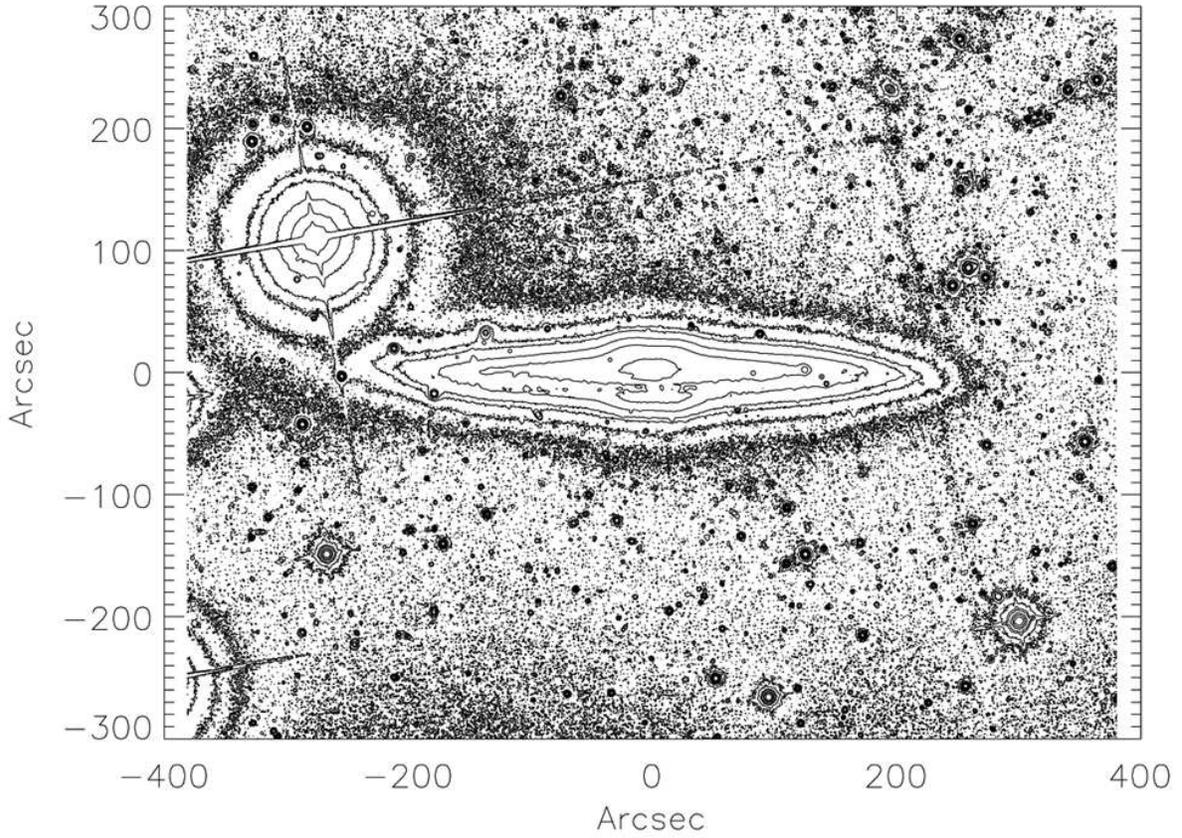}
\caption{A brightness contour plot of NGC 5746 made from the 8100 s $R$-band image of NGC 5746 in Figure~
\ref{cutboxes}.  The contours are at -9, 10, 30, 75, 125, 325, 800, and 4000 ADU above the mean sky 
background as measured $\sim$10$\arcmin$ from the center of the galaxy along its minor axis.  The isophotes 
nearest the sky level do not show any obvious indication of an extended halo.  The bright star at upper left is HD 
129827.}
\label{halo}
\end{figure}

We considered the possibility that our observations might reach a limiting magnitude sufficient to reveal the 
presence of an extended halo around NGC 5746, thereby informing the choice of the functional form of the fit to 
the largest radii in our profiles.  We searched for evidence of an extended halo in a deep (8100 s) $R$-band 
image of NGC 5746 obtained with the WIYN 0.9m telescope and S2KB CCD camera at Kitt Peak National 
Observatory over two observing seasons in 2011-12.  Brightness contours of the image are shown in Figure~\ref
{halo}. Despite contamination of the faintest isophotes from the bright nearby star HD 129827, we do not find any 
convincing detection of a halo to a limiting surface brightness of $\sim$23 mag arcsec$^{-2}$ in $K_s$.  Thus we 
do not include an explicit halo component in the light profile fits at large radii.  Specific details of the fit, including 
its functional form, are not critical to the analysis presented here.   We also did not include a component 
representing a central Seyfert nucleus or a nuclear star cluster in the fit at the smallest radii.  A bright central 
point was seen in our NGC 4565 data but we did not attempt to fit it, whereas we do not see a corresponding 
point in the NGC 5746 images. 

We subtracted the disk exponential and box S\'{e}rsic fits from a given profile, leaving the profile of the 
pseudobulge itself.  Once the pseudobulge light profiles were obtained for both axes, the total luminosity of the 
pseudobulge was obtained by integrating the light in each direction and adding the results.  A proper 
comparison of light profiles of the minor and major axes of an edge-on galaxy with a boxy pseudobulge should 
take into account the fact that the box typically has an axial ratio other than 1.  As a result, a feature seen at 
radius $r$ along the minor axis will be seen at radius ($b$/$a$)$r$ along the major axis, where $a$ and $b$ are 
the sides of the box.  Our major-axis light profiles are shifted in radius by a factor of 1.4, the axial ratio of the box 
measured from the SDSS $gri$ composite image, placing the profiles on a common spatial scale.

\section{Results\label{sec:results}}

\begin{figure}
\plotone{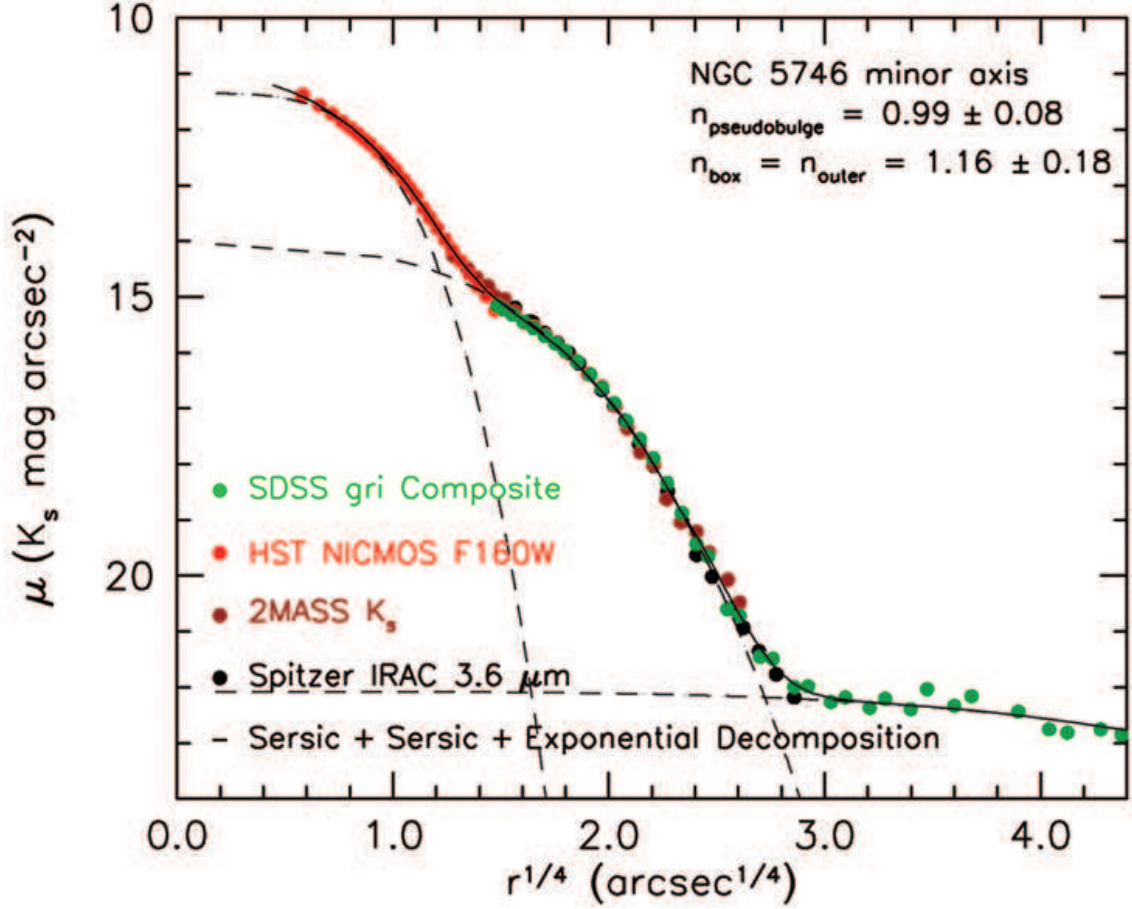}
\caption{Minor-axis light profile of NGC 5746 from the following data sources: combined SDSS $gri$ (green 
points), HST NICMOS F160W (red points), 2MASS $K_s$ (brown points), and Spitzer IRAC 3.6 $\mu$m (black 
points).  A S\'ersic-S\'ersic-exponential decomposition is overplotted as dashed lines, representing the inner 
pseudobulge, boxy bar, and outer halo, respectively.  The solid line represents the sum of these three 
components.}
\label{minor_profile}
\end{figure}

\begin{figure}
\plotone{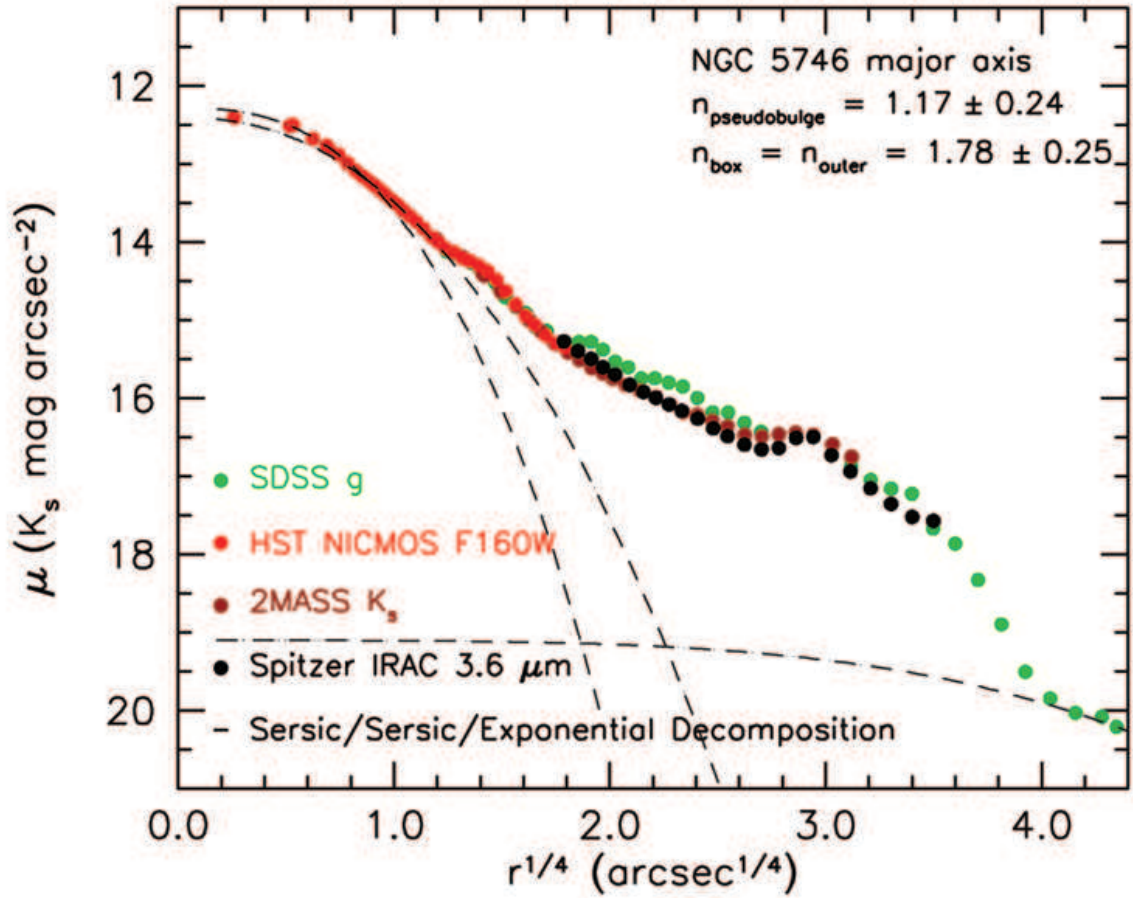}
\caption{Major-axis light profile of NGC 5746.  The data sources are the same as in Figure~\ref{minor_profile}.  A S\'ersic-S\'ersic-exponential decomposition is overplotted as dashed lines.}
\label{major_profile}
\end{figure}

The minor- and major-axis light profiles we obtain for NGC 5746 are shown in Figures~\ref{minor_profile} and~
\ref{major_profile}, respectively.  A simultaneous decomposition of the box and exponential disk yielded S\'ersic 
indices of $n$ = 1.16 $\pm$ 0.18 along the minor axis and 1.78 $\pm$ 0.25 along the major axis.  Fitting the box 
along the major axis is complicated by the presence of a bright ring, described below, seen in the infrared 
imagery.  Interior to this ring, at radii 1.8 arcsec$^{1/4}$ $\lesssim$ $r^{1/4}$ $\lesssim$ 2.7 arcsec$^{1/4}$, the 
infrared flux is low relative to the optical flux.  

The central pseudobulge was fitted with a single S\'ersic component over only the inner $\sim$2$\arcsec$, 
corresponding roughly to the angular extent of the NICMOS data.  The S\'ersic indices of the central 
pseudobulge are 0.99 $\pm$ 0.08 along the minor axis and $n$ = 1.17 $\pm$ 0.24 along the major axis.  Prior 
evidence of a bar in NGC 5746 \citep{BureauFreeman99} is bolstered by our finding of $n$ $<$ 2 for the box.  
Therefore, the true bulge in NGC 5746 is a pseudobulge, consistent with both our results for NGC 4565 and the 
observational definition in \citet{FisherDrory08} that bulges with $n$$\ltsim$2 are pseudobulges, not classical 
bulges.   It is also evident that, as in NGC 4565, the apparent ``boxy bulge'' of NGC 5746 is in fact a bar seen 
nearly end-on.  The emerging picture of NGC 5746 is that viewed face on, it would have a bar and an inner ring 
much like those inferred in NGC 4565.

\begin{figure}
\plotone{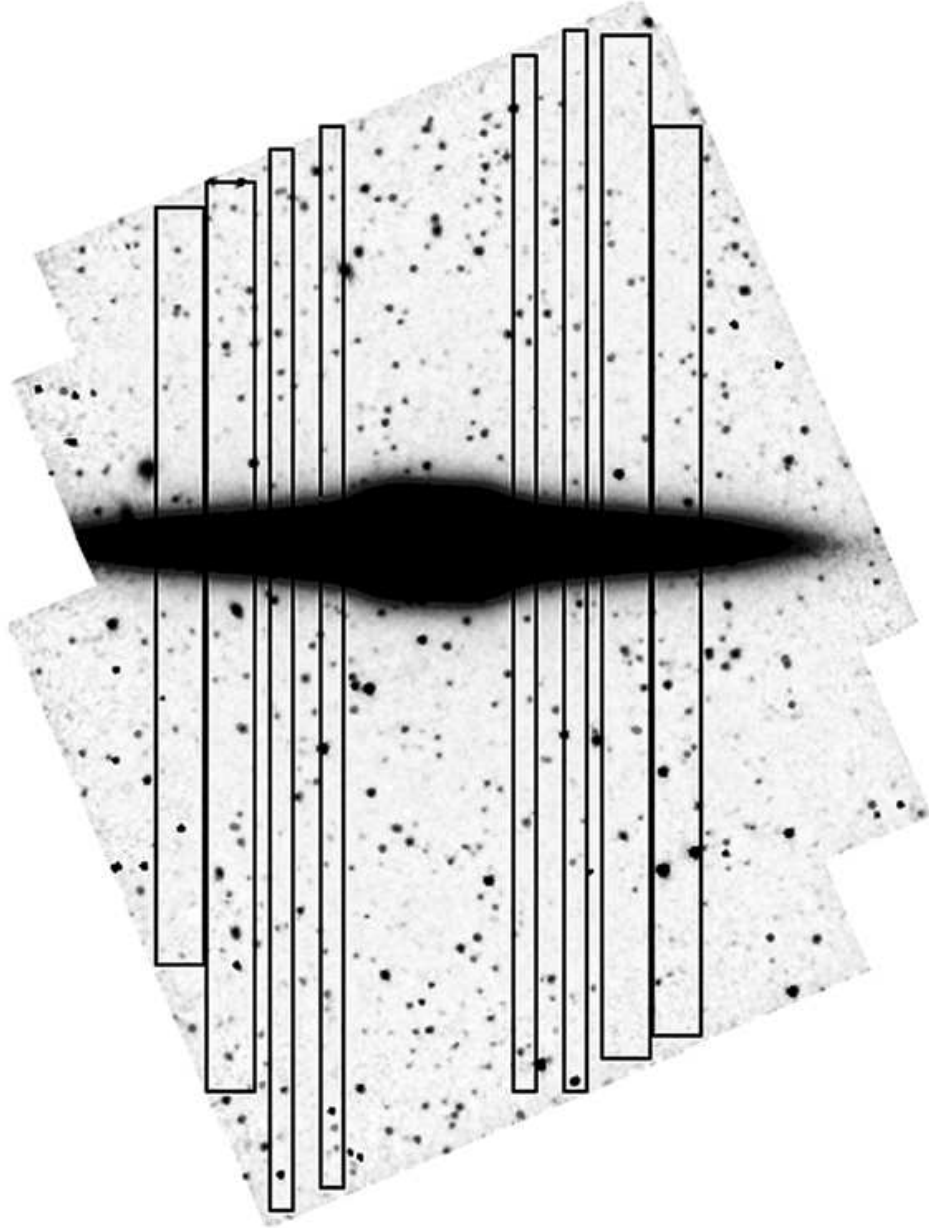}
\caption{The IRAC 3.6 $\mu$m image of NGC 5746, rendered with an inverted colormap, showing the 
dimensions and extent of the cut boxes used to extract minor axis light profiles for determination of the thick disk 
scale height as a function of radial distance along the major axis.  The image has been rotated such that the 
major axis is aligned with the image rows.}
\label{thickcuts}
\end{figure}

The fits give additional information about the nature of the central pseudobulge.  The best S\'ersic fits yield a 
scale height of 0$\farcs$74 $\pm$ 0$\farcs$10 along the minor axis and a scale length of 0$\farcs$64 $\pm$ 0$
\farcs$20 along the major axis.  At our adopted distance to NGC 5746, these correspond to 100 $\pm$ 13 pc 
and 86 $\pm$ 27 pc, respectively.   From the fit to the boxy bar plus the disk, we compute a scale height of 755 $
\pm$ 145 pc.  We measured the thick disk scale height as a function of radial distance along the major axis from 
the IRAC 3.6 $\mu$m data; the cuts we used are shown superimposed on the IRAC image in Figure~\ref
{thickcuts}.  Again, the cut box widths vary to preserve resolution at small radii along the major axis and $S$/$N$ 
at larger radii.  The mean of eight measurements of the thick disk scale height at $r$ $>$ 38$\arcsec$ (51 kpc) is 
8$\farcs$6 $\pm$ 0$\farcs$7 (1.2 $\pm$ 0.1 kpc).  For comparison, we found for NGC 4565 a pseudobulge scale 
height of 90 pc, a boxy bar plus disk scale height of 740 pc, and a thick disk scale height of 1.03 kpc.

We integrate the fits to the various components of the light profiles and find a mean value and 1$\sigma$ scatter 
of 0.136 $\pm$ 0.019 for the pseudobulge-to-total ratio ($PB/T$) of NGC 5746.  In Table~\ref{summary}, we 
show the S\'ersic $n$ values measured along both minor and major axes and the resulting $PB/T$ ratio for this 
galaxy along with those reported for NGC 4565 in \citet{KB10} for purposes of comparison.  Measurements of the 
index of the boxy component of the apparent bulge and the true (pseudo)bulge are given along with the 
consequent pseudobulge-to-total ($PB/T$) ratios.  The pseudobulge in NGC 5746 contributes relatively little 
light to the galaxy.  \citet{Balcells07}, following a similar procedure, quote a S\'ersic index of $n$ = 1.55 $\pm$ 
0.14 and a $B$/$D$ ratio of 0.1 for this galaxy.  While we recognize the care with which their measurements 
were made, we note that in neither case was allowance made for a nuclear component which could introduce 
uncertainty into values of the S\'ersic $n$.  However, the differences are not enough to affect our fundamental 
conclusion, one which the measurements of Balcells et al. support -- the bulge in NGC 5746 is a pseudobulge.  

\begin{figure}
\plotone{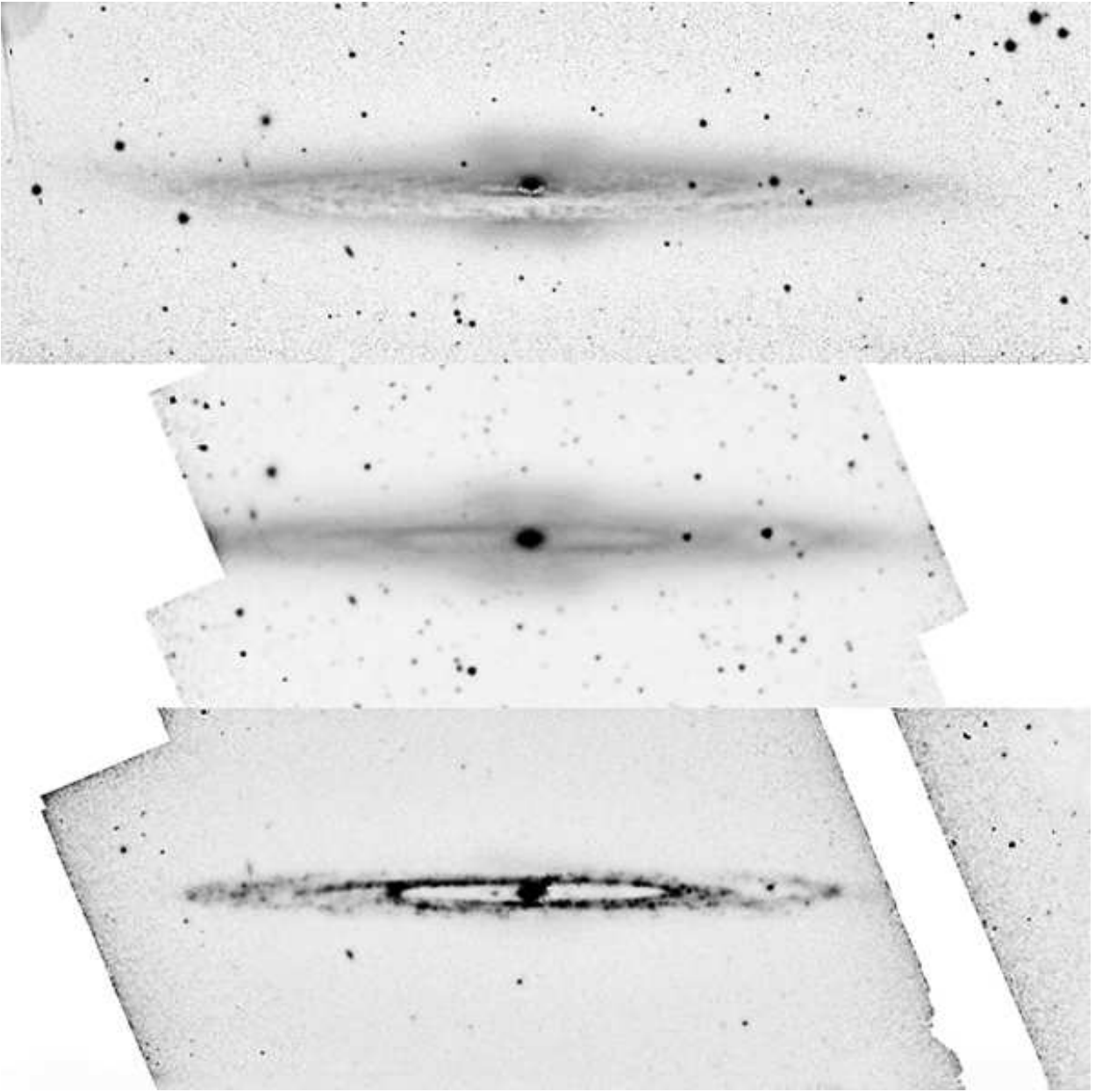}
\caption{Three views of NGC 5746 from the optical to the mid-infrared: the sum of SDSS $gri$ (top), \it Spitzer 
\rm IRAC 3.6 $\mu$m (middle), and IRAC 8 $\mu$m (bottom).  The images have been rotated such that the 
major axis is aligned with the image rows.  Irregularities in the IRAC images are caused by 
boundaries of the mosaicked regions used to make the composite image in each case.}
\label{3views}
\end{figure}

Figure~\ref{3views} shows the 3.6 $\mu$m and 8 $\mu$m IRAC images along with the SDSS $gri$ composite for 
optical context, each individually stretched to emphasize the inner ring.  Inspection of the 3.6 $\mu$m image 
clearly indicates a bright ring with an inner radius of $\sim$57$\arcsec$ (9.1 kpc) and an average radial width of 
10$\arcsec$ (1.6 kpc) as measured at the tangent points.  The ring is clearly present in Figure~\ref{major_profile} 
as a shallow rise in the major-axis light profile between 2.7 arcsec$^{1/4}$ $\leq$ $r^{1/4}$ $\leq$ 3.2 arcsec$^
{1/4}$ and is slightly brighter at its outer edge than inner edge, particularly at optical wavelengths.  In the IRAC 8 
$\mu$m band, associated with emission by polycyclic aromatic hydrocarbons (PAHs; \citealt{PAH84,PAH85,
PAH89,PAH892,PAH99,PAH00}), the ring is considerably brighter, indicating a high rate of star formation 
\citep{Verstraete01,Peeters04, Wu05, Calzetti07, Bendo08}.

\begin{figure}
\plotone{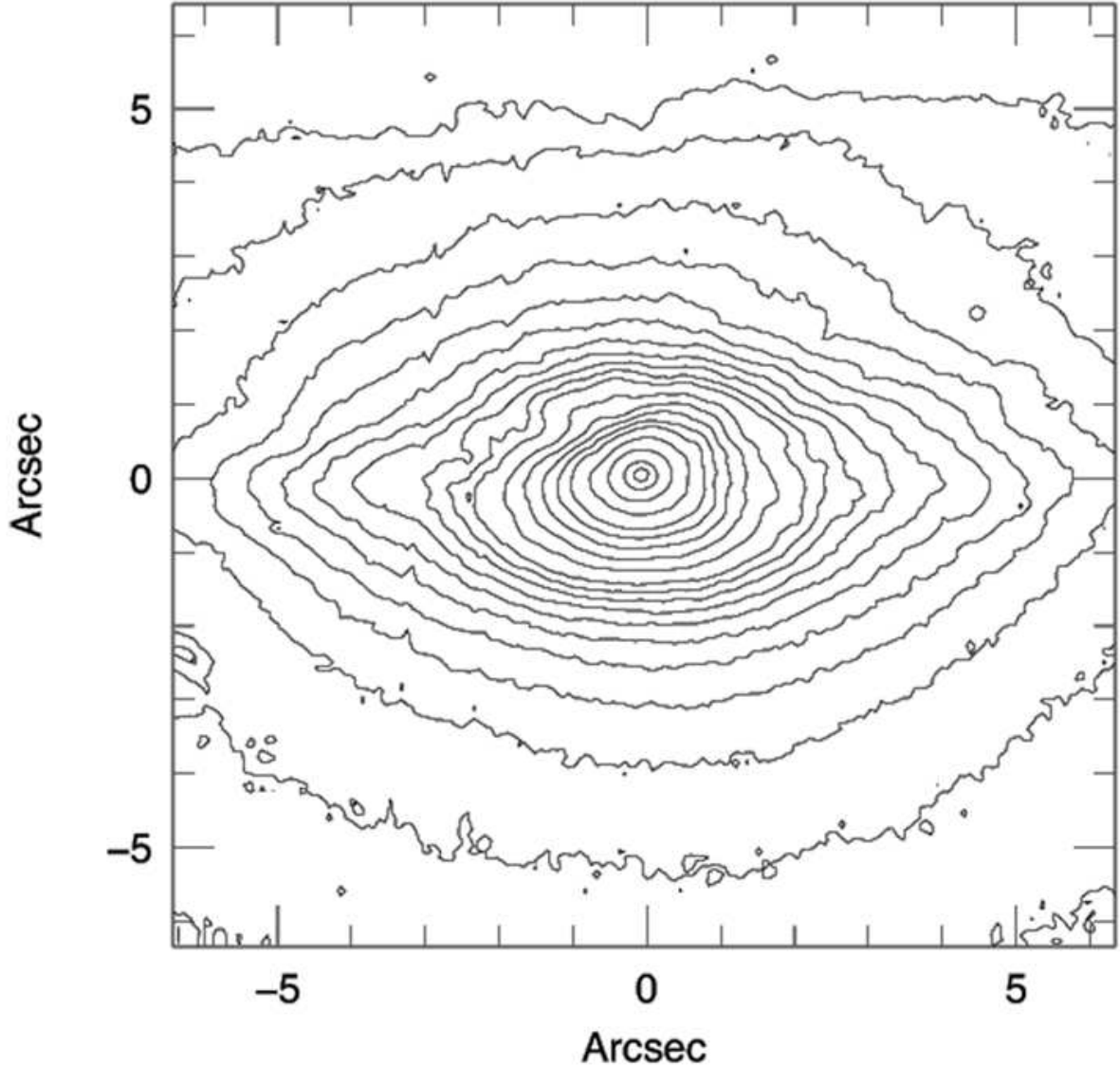}
\caption{A brightness contour plot of the nuclear region of the NICMOS $H$-band image of NGC 5746.  The 
image has been derotated such that the galaxy's major axis lies along the rows of the diagram and was cleaned 
of contaminants before the contours were generated.  The contours are at pixel levels of 0.1, 0.5, 1, 1.5, 2, 2.5, 3, 
3.5, 4, 4.5, 5, 6, 7, 8, 9, 10, 12, 15, 20, and 25 ADU after sky subtraction.  The vertical and horizontal scales are 
arranged such that the coordinate origin corresponds to the highest pixel value in the frame.  Deformation of the 
isophotes in the upper-left quadrant is due to absorption by the dust lane.}
\label{contours}
\end{figure}

Another slight rise in the major-axis profile exists at much smaller radii, from 1.2 arcsec$^{1/4}$ $\leq$ $r$$^
{1/4}$ $\leq$ 1.6 arcsec$^{1/4}$ (280 pc $\leq$ $r$ $\leq$ 880 pc).  This feature appears to be real as three of 
the four data sources trace it; the 2MASS data are not useful at small radii and therefore do not cover the radius 
range of this shallower feature.  Brightness contours of the inner 12$\farcs$9 $\times$ 12$\farcs$9 (1.7 kpc $
\times$ 1.7 kpc) of the NICMOS image are shown in Figure~\ref{contours}; the isophotes transition from elliptical 
to disky over the radius range indicated by the bump in the major-axis light profile.  This may indicate the 
presence of a nuclear disk partially obscured by dust.  In the region defined by elliptical isophotes, the profile is 
nearly linear in $r^{1/4}$ until it turns over at the limit of the NICMOS resolution.  It is well fitted by a S\'ersic 
function with an index very similar to that observed along the major axis of NGC 4565.  We believe this is 
additional evidence indicating the presence of a central pseudobulge.

The major-axis light profile is consistently brighter in the SDSS optical colors between the inner bright feature 
and the ring over an approximate radius range of 1.8 arcsec$^{1/4}$ $\leq$ $r$$^{1/4}$ $\leq$ 2.7 arcsec$^
{1/4}$ (1.4 kpc $\lesssim$ $r$ $\lesssim$ 7.1 kpc).  The bluer light again dominates from 3.2 arcsec$^{1/4}$ $
\leq$ $r$$^{1/4}$ $\leq$ 3.4 arcsec$^{1/4}$ (14.0 kpc $\lesssim$ $r$ $\lesssim$ 17.9 kpc).  This is roughly the 
radius range between the outer edge of the ring and what in the 24 $\mu$m IRAC image appears to be the inner 
edge of a set of spiral arms.  In both cases, interior to the ring and between the ring and spiral arms, the apparent 
blue color excess in the light profile results from starlight evenly distributed in the plane of the disk.  PAH 
emission in the inner ring contributes as much flux in the near- and mid-IR as the starlight does, whereas in the 
case of the spiral arms, there are insufficient IR data at large radii to determine the relative contribution of stars 
and PAH emission.  The last IRAC data point in Figure~\ref{major_profile} suggests the profile is rising in the 
mid-IR as it crosses the spiral arms and would again dominate in this region if data existed.  Seeing relatively 
unextincted starlight in the region between the inner ring and the bar is consistent with removal of gas and dust 
from this region, and should be observable in other inner-ring galaxies.

\section{Summary\label{sec:summary}}

Using archival data from multiple ground-based surveys and spacecraft missions, we carried out surface 
photometry on the almost-edge-on spiral galaxy NGC 5746 and extracted 1D light profiles along the galaxy's 
minor and major axes.  The profiles were decomposed into multiple S\'{e}rsic and exponential functions 
corresponding to the central pseudobulge, the boxy bar, and the disk.  We computed S\'{e}rsic indices for the 
pseudobulge, finding $n$ = 0.99 $\pm$ 0.08 and 1.17 $\pm$ 0.24 for the minor and major axes, respectively, 
and an axial ratio of $\sim$0.85.  The degree of flattening of this pseudobulge is consistent with the amount we 
found previously for the pseudobulge in NGC 4565.  We further note that the central pseudobulge is the most 
compact component of the light profiles, with a scale height of only 100 $\pm$ 20 pc.   Near- and mid-IR imagery 
reveals the presence of an inner ring of inner radius 9.1 kpc and width of 1.6 kpc, another structure whose 
presence is consistent with secular evolution.  The ring and a set of outer spiral arms are clearly indicated in the 
major axis light profile as well.  All this implies that NGC 5746 is a well developed, nearly-edge-on SB(r)bc 
galaxy. 

Thus NGC 5746, like NGC 4565, is a giant galaxy whose structure shows no recognizable remnant of a
recent major merger.  Such galaxies are common in field environments.  It remains difficult to understand
how they form in a hierarchically clustering universe.

\acknowledgements 
{
These results are based on observations made with the Spitzer Space Telescope, which is operated by the Jet 
Propulsion Laboratory, California Institute of Technology under a contract with NASA.  Additional data from the 
NASA/ESA Hubble Space Telescope were used, obtained from the data archive at the Space Telescope 
Institute, operated by the association of Universities for Research in Astronomy, Inc. under the NASA contract 
NAS 5-26555.  This publication makes use of data products from the Two Micron All Sky Survey, which is a joint 
project of the University of Massachusetts and the Infrared Processing and Analysis Center/California Institute of 
Technology, funded by the National Aeronautics and Space Administration and the National Science 
Foundation.  Funding for the SDSS and SDSS-II has been provided by the Alfred P. Sloan Foundation, the 
Participating Institutions, the National Science Foundation, the U.S. Department of Energy, the National 
Aeronautics and Space Administration, the Japanese Monbukagakusho, the Max Planck Society, and the Higher 
Education Funding Council for England. The SDSS Web Site is http://www.sdss.org/.  The SDSS is managed by 
the Astrophysical Research Consortium for the Participating Institutions. The Participating Institutions are the 
American Museum of Natural History, Astrophysical Institute Potsdam, University of Basel, University of 
Cambridge, Case Western Reserve University, University of Chicago, Drexel University, Fermilab, the Institute 
for Advanced Study, the Japan Participation Group, Johns Hopkins University, the Joint Institute for Nuclear 
Astrophysics, the Kavli Institute for Particle Astrophysics and Cosmology, the Korean Scientist Group, the 
Chinese Academy of Sciences (LAMOST), Los Alamos National Laboratory, the Max-Planck-Institute for 
Astronomy (MPIA), the Max-Planck-Institute for Astrophysics (MPA), New Mexico State University, Ohio State 
University, University of Pittsburgh, University of Portsmouth, Princeton University, the United States Naval 
Observatory, and the University of Washington.  

This work was supported by the National Science Foundation 
under grant AST-0607490 and by the Curtis T.~Vaughan, Jr.~Centennial Chair in Astronomy at the University of
Texas at Austin.
}

\bibliographystyle{apj}
\bibliography{barentine}

\clearpage

% Tables 

\begin{deluxetable}{ccccc}
\tabletypesize{\footnotesize}
\tablecolumns{5}
\tablewidth{0pt}
\tablecaption{Properties of Data Sources}
\tablehead{ \colhead{Telescope} & \colhead{Instrument} & \colhead{Field Of View (\arcmin)} & \colhead{Scale ($\arcsec$ pix$^{-1}$)} & \colhead{Filters} }
\startdata
Spitzer 0.85m & IRAC & 5.22 $\times$ 5.22 & 1.2 $\times$ 1.2 & Channel 1 (3.6$\mu$m) \\
HST 2.4m & NICMOS & 0.3 $\times$ 0.3 & 0.076 $\times$ 0.075 & F160W (1.6$\mu$m) \\
2MASS 1.3m & 2MASS Camera & 8.5 $\times$ 8.5 & 2.0 $\times$ 2.0 & $K_s$ \\
SDSS 2.5m & Imager & 13.51 $\times$ 8.98\tablenotemark{a} & 0.4 $\times$ 0.4 & $g, r, i$
\enddata
\tablenotetext{a}{Dimensions are given for a single image frame.}
\label{instruments}
\end{deluxetable} 

\begin{deluxetable}{lccccccc}
\tabletypesize{\footnotesize}
\tablecolumns{7}
\tablewidth{0pt}
\tablecaption{S\'ersic Indices and Pseudobulge-To-Total (PB/T) Ratios For NGC 4565 and NGC 5746}
\tablehead{\colhead{Designation} & \colhead{$v_c$\tablenotemark{a} (km s$^{-1}$)} &
	  \multicolumn{2}{c}{Box S\'ersic $n$} & 
           \multicolumn{2}{c}{Pseudobulge S\'ersic $n$} &
           \colhead{PB/T} \\ &
           \colhead{} &
           \colhead{Minor} & 
           \colhead{Major} &
           \colhead{Minor} &
           \colhead{Major} &
           \colhead{}
}
\startdata
NGC 4565 & 255$\pm$10\tablenotemark{b}     & 1                          & --                          & 1.33$\pm$0.12 & 1.55$\pm$0.07 & 0.061$\pm$0.010 \\
NGC 5746 & 318.5$\pm$9.8\tablenotemark{c} & 1.16$\pm$0.18 & 1.78$\pm$0.25 & 0.99$\pm$0.08 & 1.17$\pm$0.24 & 0.136$\pm$0.019
\enddata
\tablenotetext{a}{Maximum circular velocity, corrected for the inclination angle.}
\tablenotetext{b}{\citet{Rupen91}}
\tablenotetext{c}{\citet{Kuijken95}}
\label{summary}
\end{deluxetable}

% Figures

\end{document}